\documentclass[dvips,11pt]{article}

\usepackage{verbatim}
\usepackage{graphicx}
\usepackage{amssymb,amscd}
\usepackage{amsmath}
\usepackage{bbm}
\usepackage{epsfig}
\usepackage{epsf}
\usepackage{pslatex}
\usepackage{colortbl}
\usepackage[margin=2.0cm]{geometry}

\newtheorem{thm}{Theorem}[section]

\numberwithin{equation}{section}

\newtheorem{algorithm}[thm]{Algorithm}
\newcommand{\lam}{\lambda}
\newcommand{\be}{\begin{equation}}
\newcommand{\ee}{\end{equation}}
\begin{document}

{\bf \title { Adjoint method for a tumour growth PDE-constrained optimization problem.}}

\author{ D. A. Knopoff, D. R. Fern\'andez, G. A. Torres and C. V. Turner \thanks{E-mail address:  knopoff@famaf.unc.edu.ar, dfernandez@famaf.unc.edu.ar, torres@gmail.com, turner@famaf.unc.edu.ar} }
\date{}

\maketitle

\begin{center}
\small{ {\it FaMAF, Universidad Nacional de C\'ordoba -
CIEM-CONICET. C\'ordoba, Argentina.}}
\end{center}

\begin{abstract}

 In this paper we present a method for estimating unknown parameters that appear on an avascular, spheric tumour growth model.
  The model for the tumour is based on nutrient
driven growth of a continuum of live cells, whose birth and death
generate volume changes described by a velocity field. 

The model consists on a coupled system of partial differential
equations whose spatial domain is the tumour, that changes in size over 
time. Thus, the situation can be formulated as a free boundary problem. 

After solving the forward problem properly, we use the model for the 
estimation of parameters by fitting the numerical solution with real data,
obtained via in vitro experiments and medical imaging. We define an appropriate functional to compare both the real data and the numerical solution.
We use the adjoint method for the minimization of this functional, getting a better performance than the obtained with the pattern search method.
\end{abstract}


\maketitle

\section{Introduction.}
The interest for research in modeling cancer has grown enormously
over the last decades \cite{Adam1,Adambellomo,BeChaDe09,BellomoLiMaini}, and it became one of the most challenging topics involving applied mathematicians working with researchers in the biological sciences. One of the main motivations is the fact that, according to the World Health Organization, about six million people die annually because of cancer, being this one the second main fatal disease in the industrial countries. 

Key comments on the importance of mathematical modeling in cancer can be found in a vast part of the literature. For example, in the work by Bellomo \textit{et al.} \cite{BellomoLiMaini}, they emphasize the fact that ``applied mathematics may be able to provide a framework in which experimental results can be interpreted, and a quantitative analysis of external actions to control neoplastic growth can be developed''. Moreover, ``models and simulations can reduce the amount of experimentation necessary for drug and therapy development''.

In this paper we consider the case of avascular multicellular spheroids
(MCS). Pioneers in this subject have
been, for example \cite{Greenspan, Shymko}, where the first
spatio-temporal models of MCS' growth have been developed. The study of MCS is interesting because they provide the best insight into the effects of varying nutrient concentrations or the effectiveness of
chemotherapeutic drugs on tumours in vivo, and their behaviour can
be studied experimentally (in vitro) by controlling environmental
conditions in which they grow: for example, the radii of the tumour
can be monitored while changing the chemotherapeutic drug or oxygen
levels.

In addition, another variables can be measured. If possible,
experimentalists can get information about the distribution of
substances within the tumour. Moreover, via medical imaging, 
histopathology and potentially other sources, they can also get data about the density of the
different kind of cells conforming it: proliferating, quiescent,
necrotic. For instance, as documented in \cite{cnea} the Boron Neutron Capture Therapy (BNCT) technique gives information about the evolution in size of a melanoma, and in \cite{hogea}, they obtain information about the growth of a glioma via Magnetic Resonance Imaging (MRI).

That is why in this general approach of modeling, the key variables
are the tumour size (radius), the concentration within the tumour
of growth-rate limiting diffusible chemicals (nutrients such as
oxygen or glucose or a chemotherapeutic drug) and the density of cells. Since the tumour
changes in size over time, the domain on which the models are
formulated must be determined as a part of the solution process,
giving a vast class of moving boundary problems \cite{Byrne2,
Crank}.

In this article, we propose a framework for estimating unknown
parameters via a PDE-constrained optimization problem, following the PDE-based
model by Ward and King \cite{WK2}, and considered also by Knopoff \textit{et al.} \cite{knopoff}. In this approach, avascular
tumour growth is modeled via a coupled nonlinear system of partial
differential equations, which makes the numerical solution procedure
quite challenging.

This kind of problem constitutes a particular application of the so-called inverse problems, which are being increasingly used in a broad number of fields in applied sciences. For instance, problems refered to structured population dynamics \cite{PeZu07}, computerized tomography and image reconstruction in medical imaging \cite{DoAsPa11, ZuMa03}, and more specifically tumour growth \cite{AgBaTu, hogea, knopoff}, among many others.

Extending the work done in \cite{knopoff}, we are concerned with developing a robust PDE-constrained
formulation that let us find the best set of parameters of a tumour
growth model that fits patient or experimental data. In contrast to the previous work, in which we used a free-derivative method (pattern search algorithm) we now use the adjoint method in order to find the derivative of the afore mentioned functional. We want to find the
parameters that would be of interest by defining a functional to be minimized. In this way, we would obtain the best set of parameters that fits patient-specific data.

The contents of this paper, which is organized into 9 sections and an appendix, are as follows: Section 2 consists in some preliminaries about the model and the definition of the direct problem. It can be considered as a revision of our previous work. Section 3 deals with the motivation and formulation of the minimization problem. Section 4 introduces the adjoint problem, deriving the optimality conditions for the problem. Specifically, we show how the adjoint method may be used to find the derivative of the solution of a PDE with respect to a parameter that does not appear explicitly in the equation. Section 5 refers to specification of the tools used in Section 4 for the concrete problem. Section 6 deals with the numerical solution of the adjoint problem, designing a suitable algorithm to solve it. In particular, we develop a method to deal with some singularities in the PDEs. Section 7 deals with the minimization method to be used and, specifically, we propose an algorithm to find a minimum of the functional, by choosing a proper descent direction using derivatives. In section 8 we show some numerical simulations to give information on the behaviour of the function and its dependence on the parameters. Section 9 presents the conclusions and introduces some future work related to the contents of this paper.

Some words about our notation.
We use $\langle\cdot,\cdot\rangle$ to denote the $L^2$ inner product (the space is always clear from the context) 
and we consider the sum of inner products for a cartesian product of spaces.
For a function $F:\mathcal{Y}\times {\mathcal U}\rightarrow \mathcal{Z}$ such that $(\phi,p)\mapsto F(\phi,p)$, we denote by
$F'(\phi,p)$ the full Fr\'echet-derivative and by $\frac{\partial F}{\partial \phi}(\phi,p)$ and $\frac{\partial F}{\partial p}(\phi,p)$
the partial Fr\'echet-derivatives of $F$ at $(\phi,p)$. For a linear operator $T : \mathcal{Y} \rightarrow \mathcal{Z}$ we denote $T^* : \mathcal{Z}^* \rightarrow \mathcal{Y}^*$ the adjoint operator of $T$. If $T$ is invertible, we call $T^{-*}$ the inverse of the adjoint operator $T^*$.

\section{Some preliminaries about the model.}

In \cite{knopoff} we dealt with the resolution of a coupled system of spatio-temporal PDEs which involves initial and boundary conditions, with the additional difficulty that the boundary is also an unknown. In that work we considered that the
tumour is a spheroid which consists of a continuum of living cells,
in one of two states: live or dead. The rates of birth and death
depend on the nutrient and chemotherapeutic drug concentration. It
is supposed that those processes generate volume changes, leading to
cell movement described by a velocity field. In order to develop a method capable to recover parameters, we will consider the simpler case in which no treatment is supplied to the neoplastic formation, leaving this case for future work. Under this assumption, the system of equations
to be studied is:

\begin{eqnarray}
\frac{\partial \eta}{\partial t}+\frac{1}{r^2}\frac{\partial(r^2 \nu\eta)}{\partial r}&=&[k_m(\varsigma,\theta)-k_d(\varsigma,\theta)]\eta, \label{eq1}
\\
\frac{\partial \varsigma}{\partial
t}+\frac{1}{r^2}\frac{\partial(r^2 \nu\varsigma)}{\partial
r}&=&\frac{D}{r^2}\frac{\partial}{\partial r}\left(r^2\frac{\partial
\varsigma}{\partial r}\right)-\beta k_m(\varsigma,\theta)\eta, \label{eq2} 
\\
 \frac{1}{r^2}\frac{\partial(r^2 \nu)}{\partial r}&=&[V_L
k_m(\varsigma,\theta)-(V_L-V_D)k_d(\varsigma,\theta)]\eta, \label{eq3}
\end{eqnarray}
where the dependent variables $\eta$, $\varsigma$ and $\nu$  are the live
cell density (cells/unit volume), nutrient concentration and velocity, respectively. The independent variables are the radial position $r$ inside the tumour and time $t$. Constants $V_L$ and $V_D$ correspond to the volume of a living and a death cell, respectively. The number $D$ is the diffusion coefficient of the nutrient and $\beta$ is a positive constant related to the nutrient's consumption rate. As it is described in
\cite{WK2}, equation (\ref{eq1}) states that the rate of change of
$\eta$ is dependent on the difference between the birth $k_m(\varsigma,\theta)$ and
death $k_d(\varsigma,\theta)$ rates ($\theta$ is a vector of parameters associated to these functions). The functions $k_m$ and $k_d$ are taken to be generalized
Michaelis-Menten kinetics with exponent 1, i.e.

\begin{eqnarray}
k_m(\varsigma,\theta)&=&A\left(\frac{\varsigma}{\varsigma_c+\varsigma}\right), \label{mm1}
\\
k_d(\varsigma,\theta)&=&B\left(1-\sigma\frac{\varsigma}{\varsigma_d+\varsigma}\right), \label{mm2}
\end{eqnarray}
where $\theta = [A, B, \varsigma_c, \varsigma_d, \sigma]^T$ are model parameters. As stated in the appendix of \cite{WK1},there appears to be no appropriate data available on the parameters $\varsigma_d$ and $\sigma$, constituting one extra motivation for this work. 

Inherent in this problem are two timescales: the tumour growth timescale ($\approx$ 1 day) and the much shorter nutrient diffusion ($\approx$ 1 min), letting us to adopt a quasisteady assumption in the nutrient equation (see \cite{WK1}). Therefore, we replace (\ref{eq2}) by the quasisteady approximation
\begin{equation}\label{eq2_QS}
\frac{1}{r^2}\frac{\partial}{\partial r}\left(r^2\frac{\partial
\varsigma}{\partial r}\right)=\frac{\beta}{D} k_m(\varsigma,\theta)\eta. 
\end{equation}

\subsection{Initial and boundary conditions.}

As it has been mentioned, the tumour is assumed to be a spheroid
that exhibits radial symmetry. That is why, not only the state
variables $\eta$, $\varsigma$ and $\nu$ are important, but also the tumour
radius is a key variable to be determined. Since the tumour
changes in size over time, the domain on which the model is
formulated (and the PDEs are valid) must be determined as part of
the solution.

Let $\mathcal{S}(t)$ be the tumour radius at time $t$. At $t = 0$ we will consider the tumour at a certain stage of its evolution. Hence the initial conditions are a known radius $\mathcal{S}(0)$ and an initial live cell density
\[
\eta(r,0)  =  \eta_I(r). 
\]
Because symmetry is assumed about the tumour center, there is no
flux there. That is why the boundary conditions at $r=0$ are:
\begin{eqnarray}
\label{bc1}
\frac{\partial \varsigma}{\partial r}(0,t)&=&0, \\ 
\nu(0,t)&=&0.
\end{eqnarray}

Moreover, on the external boundary (which is also the boundary of
the complement of the tumour as a subset of the body), the following
conditions are taken:
\begin{eqnarray}
\label{bc2}
 \varsigma(\mathcal{S}(t),t)&=&c_0, \\ 
 \frac{d\mathcal{S}}{dt}&=&\nu(\mathcal{S}(t),t),
\end{eqnarray}
where $c_0$ is the external nutrient concentration.

\subsection{Nondimensionalization and fixed domain method.}

Following the ideas exposed in \cite{Adam1,Adambellomo,Byrne2,WK1,WK2}, the mathematical model is rescaled and the domain $[0, \mathcal{S}(t)]$ of the tumour is transformed onto the interval $[0,1]$. This is a very useful approach when dealing with free-boundary problems, as mentioned in \cite{Crank}. Hence, let us define the following functions
\begin{eqnarray*}
 N(y,t)&=& V_L \eta(y\mathcal{S}(t/A),t/A),\\
 C(y,t)&=& \frac{1}{c_0} \varsigma(y\mathcal{S}(t/A),t/A), \\
 V(y,t)&=& \frac{1}{Ar_0}\nu(y\mathcal{S}(t/A),t/A),\\
 S(t)&=&\frac{1}{r_0}\mathcal{S}(t/A),\\
a(c,\vartheta)&=&\frac{1}{A}[k_m(c,\vartheta)-k_d(c,\vartheta)],\\
b(c,\vartheta)&=&\frac{1}{A}[k_m(c,\vartheta)-(1-\delta)k_d(c,\vartheta)],\\
k(c,\vartheta)&=&\widehat{\beta} k_m(c,\vartheta),
\end{eqnarray*}
where $r_0=(3V_L/4 \pi )^{1/3}$ is the radius of a single cell, $\delta=V_D/V_L$, $\widehat\beta=r_0^2 \beta/(V_L c_0 D)$ and $\vartheta= [A,B,c_c,c_d,\sigma]$ with $c_c=\varsigma_c/c_0$ and $c_d=\varsigma_d/c_0$. Thus, we obtain the following system to be solved:
\begin{eqnarray}
\label{eqN}
N_t-\frac{S^{\prime}}{S}yN_y + \frac{V}{S}N_y &=& N[a(C,\vartheta)-b(C,\vartheta)N], \quad
0<y\leq 1, \quad t>0,\\
\label{eqC}
C_{yy}+\frac{2}{y}C_y &=& k(C,\vartheta)S^2N, \quad
0<y\leq 1, \quad t>0,\\
\label{eqV}
V_y+\frac{2}{y}V &=& b(C,\vartheta)NS,\quad
0<y\leq 1, \quad t>0.
\end{eqnarray}

The initial conditions for the transformed problem are:
\begin{eqnarray}
N(y,0)&=&N_I(y), \quad 0\leq y\leq 1, \\
S(0)&=&S_I, 
\end{eqnarray}
where $ N_I(y)= V_L \eta_I(y\mathcal{S}(0),0)$ and $S_I = \mathcal{S}(0)/r_0$, and the boundary conditions are:
\begin{eqnarray}
V(0,t)&=&0, \quad t>0, \\
C_y(0,t)&=&0, \quad t>0, \\
C(1,t)&=&1, \quad t>0, \\
 S^{\prime}(t)&=&V(1,t), \quad t>0  \label{Scondition}.
\end{eqnarray}

From now on, equations (\ref{eqN})-(\ref{Scondition}) will be referred to as the direct problem.

\section{Formulation of the minimization problem.}

As described above, there is a set of parameters (some of them unknown) that determines the behaviour of a tumour growth. For this reason we propose to use an inverse problem technique in order to estimate them.

We define the following vectors:
\begin{eqnarray}
\phi&=&[N, V, C, S]^T, \label{statevar}
\\
p&=&[c_c, c_d, \sigma]^T,\label{parameters}
\end{eqnarray}
where $\phi$ represents the solution of the direct problem (the components of $\phi$ are the state variables of the problem) for each choice of the vector of parameters $\vartheta=(A,B,p)$, where $A$ and $B$ are assumed to be constants. Hence from now on, we will use just $p$ instead of $\vartheta$ as the vector of parameters.

Let us assume that experimental information is available during the time interval $0\leq t \leq T$. Then, the general problem we are interested to solve can be formulated as:

\begin{center}
 \textit{Find a parameter $p$ able to
generate data $\phi=[N, C, V, S]^T$ that best match the available (experimental) information over time $0\leq t\leq T$}.
\end{center}

For this purpose, we should construct an objective functional which gives us a notion of
distance between the experimental (real) data and the
solution of the system of PDEs for each choice of parameters $p$.

\begin{figure}
\begin{center}
\includegraphics[scale=0.35]{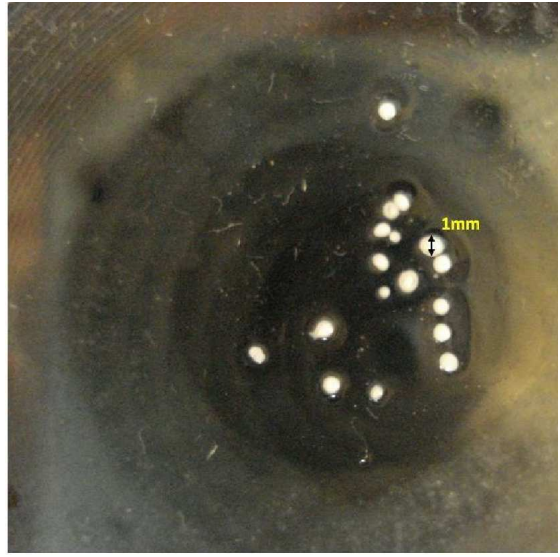}
\setlength{\belowcaptionskip}{2pt} 
\caption{Microscopic image of neoplastic colonies that grow with an external nutrient supply. \textit{Courtesy of CNEA (Comisi\'on Nacional de Energ\'ia At\'omica).} }
\label{fig:esferoides}
\end{center}
\end{figure}

First of all, it is important to decide which variables are capable
to be measured experimentally. For instance, it is clear that the
tumour radius can be known at certain times $t_k$,
$k=1,...,M$ via MRI, PET (Positron Emission Tomography) or CT (Computed Tomography). For example, Figure \ref{fig:esferoides} is a microscopic field that shows the formation \textit{in vitro} of neoplastic colonies which grow as spheroids with an external nutrient supply. Such experiments could help to determine optimal variables and parameters in order to control real tumour growth.

So, the first possibility for defining a functional could be:
\begin{equation}
\label{jota1} J(S, p)=\frac{1}{2}\int_{0}^T [S(t)-S^*(t)]^2 dt,
\end{equation}
where $S(t)$ is the radius evolution obtained by solving the
direct problem for a certain choice of $p$ and $S^*(t)= \frac{1}{r_0}\mathcal{S}^*(t/A)$ is the evolution measured experimentally (real data).

Another variable that could be measured is the density of living
cells, also via biomedical imaging. As it was mentioned before, this could be done via PET technique for a tumour in vivo, or via immunofluorescence and electronic scan microscopy technique for in vitro cases. Thus, we are motivated to define a functional that reproduces in a better way the knowledge we have about the process:

\begin{equation}\label{jota2} 
J(N,S,p)=\frac{\mu_1}{2}\int_0^1\int_{0}^T [N
(y,t)-N^*(y,t)]^2 dt dy + \frac{\mu_2}{2} \int_{0}^T [S(t)-S^*(t)]^2 dt,
\end{equation}
where $N (y,t)$ and $N^*(y,t)$ are the living cell
concentrations for the direct problem solved with the parameters $p$
and the real data, respectively (both of them in the domain
$[0,1]\times[0,T]$). The positive constants $\mu_1$ and $\mu_2$ are introduced, as
we shall see, to take into account the different order of magnitude
between $N$ and $S$. Note that, for instance, if we take $\mu_1=0$ and $\mu_2=1$ in (\ref{jota2}) we get (\ref{jota1}). In this way, these two parameters will give us some flexibility in order to choose an appropriate functional according to the experimental method used to obtain the data.

It should be noted that the spatial integration is done over the interval $[0,1]$, because we are using the solution in the fixed domain.

Let us define
\begin{equation}\label{constraint}
E\left(\phi,p\right)=\left[
\begin{array}{c}
\displaystyle N_t-N_y\frac{S^{\prime}}{S}y+\frac{V}{S}N_y-N\left(a(C,p)-b(C,p)N\right) \\[3mm]
\displaystyle V_y+\frac{2}{y}V-b(C,p)NS \\[3mm]
\displaystyle C_{yy}+\frac{2}{y}C_y-k(C,p)NS^2 \\[3mm]
\displaystyle V(1,\cdot)-S^{\prime} \\[1mm]
\displaystyle V(0,\cdot) \\[1mm]
\displaystyle C(1,\cdot)-1 \\[1mm]
\displaystyle C_y(0,\cdot) \\[1mm]
\displaystyle N(\cdot,0)-N_I \\[1mm]
\displaystyle S(0)-S_I
\end{array}
\right].
\end{equation}
In this way we can rewrite the system of PDEs (\ref{eqN})-(\ref{Scondition}) described in the previous section as $E(\phi,p)=0$.

The set of parameters that best matches the experimental data with the generated data provided by the direct problem can be computed solving a PDE constrained optimization problem, namely:
\begin{equation}
\begin{array}{rl}
\displaystyle \mathop{ \mathrm{minimize} }_p & J( \phi , p ) \\ \mathrm{subject \, to} & E(\phi,p) = 0, \\ & p \in U_{ad},
\end{array}
\end{equation}
where $U_{ad}$ denotes the set of admissible values of
$p$. In our case, according to (\ref{parameters}), $U_{ad}$ should be a subset of $\mathbbm{R}^3$. Notice that a solution $(\phi,p)$ must satisfy the constraints $E(\phi,p)=0$, which constitute the direct problem.

 We remark that, in general, there is a fundamental
difference between the direct and the inverse problems. In fact, the
latter is usually ill-posed in the sense of existence, uniqueness
and stability of the solution. This inconvenient is often treated by
using some regularization techniques \cite{DoAsPa11,engl,kirsch}. 
%	In our case we use a regularization technique that consists on a penalization of the functional $J$.  (PERO NO LO USAMOS!)

\section{Formulation of the reduced and adjoint problems.}\label{sec:adjoint}
In the following, we will consider a generic optimization problem, which has the form:
\begin{equation}\label{GP}
\begin{array}{rl}
\displaystyle \mathop{ \mathrm{minimize} }_p & J( \phi , p ) \\ \mathrm{subject \, to} & E(\phi,p) = 0, \\ & p \in U_{ad},
\end{array}
\end{equation}
where $J:\mathcal{Y}\times U_{ad}\rightarrow\mathbbm{R}$ is an objective function and $E:\mathcal{Y}\times U_{ad}\rightarrow \mathcal{Z}$ is a state equation, for $\mathcal{Y}$ and $\mathcal{Z}$ Banach spaces and $U_{ad}$ is a set of admissible points.

For completeness, in this section we will present a general theory in order to solve problem (\ref{GP}). According to the ideas exposed in \cite{branden,hinze}, we make the following assumptions:

\begin{itemize}
\item[(A1)] $U_{ad}\in\mathbbm{R}^m$ is a nonempty, closed and convex set.
\item[(A2)] $J:\mathcal{Y}\times U_{ad}\rightarrow\mathbbm{R}$ and $E:\mathcal{Y}\times U_{ad}\rightarrow \mathcal{Z}$ are continuously Fr\'echet-differentiable functions.
\item[(A3)] For each $p\in U_{ad}$ there exists a unique corresponding solution $\phi(p)\in \mathcal{Y}$ such that $E(\phi(p),p)=0$. Thus, there is a \textit{unique solution operator} $p\in U_{ad}\mapsto\phi(p)\in \mathcal{Y}$.
\item[(A4)] The derivative $\frac{\partial E}{\partial \phi} (\phi(p),p):\mathcal{Y}\rightarrow \mathcal{Z}$ is a continuous linear operator, and it is continuously invertible for all $p\in U_{ad}$.
\end{itemize}

Under these hypotheses $\phi(p)$ is continuously differentiable on $p\in U_{ad}$ by the implicit function theorem. Thus, it is reasonable to define the following so-called reduced problem

\begin{equation}
\begin{array}{rl}
\displaystyle \mathop{ \mathrm{minimize} }_p & \tilde{J}(p)= J(\phi(p),p) \\ \mathrm{subject \, to} & p \in U_{ad},
\end{array}
\end{equation}
where $\phi(p)$ is given as the solution of $E(\phi(p),p)=0$.

In order to find a minimum of the continuosly differentiable function $\tilde{J}$, it will be important to compute the derivative of this reduced objective function. Hence, we will show a procedure to obtain $\tilde{J}'$ by using the adjoint approach. Since

\begin{eqnarray*}
\left\langle \tilde{J}^{\ \prime}(p), q \right\rangle &=& \left\langle\frac{\partial J}{\partial \phi}(\phi(p),p),\phi^{\prime}(p)q\right\rangle + \left\langle \frac{\partial J}{\partial p}(\phi(p),p) , q \right\rangle
\\
& = & \left\langle \bigl( \phi^{\prime}(p) \bigl)^* \frac{\partial J}{\partial \phi} (\phi(p),p) + \frac{\partial J}{\partial p}(\phi(p),p),q \right\rangle.
\end{eqnarray*}
we see that
\begin{equation} \label{jtildedep}
\tilde{J}^{\ \prime}(p) = \bigl( \phi^{\prime}(p) \bigl)^* \frac{\partial J}{\partial \phi} (\phi(p),p) + \frac{\partial J}{\partial p}(\phi(p),p).
\end{equation}

Let us consider $\lam\in \mathcal{Z}^*$ as the solution of the so-called adjoint problem:
\begin{equation}\label{adjoint}
\frac{\partial J}{\partial \phi}(\phi(p),p)+\left(\frac{\partial E}{\partial \phi}(\phi(p),p)\right)^*\lambda=0.
\end{equation}
where $\left( \frac{\partial E}{\partial \phi}(\phi,p) \right)^*$ is the adjoint operator of $\frac{\partial E}{\partial \phi}(\phi,p)$. Note that each term in (\ref{adjoint}) is an element of the space $\mathcal{Y}^*$. 

An equation for the derivative $\phi^{\prime}(p)$ is obtained by differentiating the equation $E(\phi(p),p)=0$ with respect to $p$:
\be\label{derivE}
\frac{\partial E}{\partial \phi}(\phi(p),p)\phi^{\prime}(p)+\frac{\partial E}{\partial p}(\phi(p),p)=0,
\ee
where $0$ is the zero vector in $\mathcal{Z}$. 

By using (\ref{jtildedep}) we have that:
\begin{eqnarray*}
\tilde{J}^{\ \prime}(p) &=& \bigl( \phi^{\prime}(p) \bigl)^* \frac{\partial J}{\partial \phi} (\phi(p),p) + \frac{\partial J}{\partial p}(\phi(p),p)
\\
& = & - \left( \frac{\partial E}{\partial p}(\phi(p),p) \right)^* \left( \frac{\partial E}{\partial \phi}(\phi(p),p) \right)^{-*} \frac{\partial J}{\partial \phi}(\phi(p),p) + \frac{\partial J}{\partial p}(\phi(p),p)
\\
& = & \left( \frac{\partial E}{\partial p}(\phi(p),p) \right)^* \lambda + \frac{\partial J}{\partial p}(\phi(p),p),
\end{eqnarray*}
where in the second equation we used (\ref{derivE}) and for the last equation we used (\ref{adjoint}). Then:
\begin{equation} \label{adjoint2}
\tilde{J}^{\ \prime}(p) = \frac{\partial J}{\partial p}(\phi(p),p) + \left( \frac{\partial E}{\partial p}(\phi(p),p) \right)^* \lambda.
\end{equation}

Notice that in order to obtain $\tilde{J}^{\ \prime}(p)$ we need first to compute $\phi(p)$ by solving the direct problem, followed by the calculation of $\lambda$ by solving the adjoint problem. For computing the second term of (\ref{adjoint2}) it is not necessary to obtain the adjoint of $\frac{\partial E}{\partial p}(\phi(p),p)$ but just its action over $\lambda$ (see the Appendix).

\section{Getting the adjoint equation for the concrete problem.}

For our case, let us define $\Omega= [0,1]\times[0,T]$, the spatio-temporal domain of interest. Note that, according to (\ref{statevar}), $\phi$ is an element of a suitable vector space. Let us consider the function spaces 
\begin{eqnarray*}
& \mathcal{Y}=\left(\mathcal{C}^1(\Omega)\right)^2\times \mathcal{C}^2(\Omega)\times \mathcal{C}^1([0,T]),\\
& \mathcal{Z}=\left(\mathcal{C}^1(\Omega)\right)^2\times \mathcal{C}^2(\Omega)\times \left(\mathcal{C}^1([0,T])\right)^4 \times \mathcal{C}([0,1]) \times \mathbbm{R}.
\end{eqnarray*}
The spaces $\mathcal{C}^1$ and $\mathcal{C}^2$ inherit the inner product from $L^2$, so the spaces $\mathcal{Y}$ and $\mathcal{Z}$ are Hilbert spaces (therefore we can identify $\mathcal{Y}^*$ and $\mathcal{Z}^*$ with $\mathcal{Y}$ and $\mathcal{Z}$ respectively). It is worth mentioning that we consider these vector spaces because we look for strong solutions of the PDEs, i.e., we require differentiability of the state variables. 

In order to obtain the adjoint operator of $\frac{\partial E}{\partial \phi}$, we have to find $\left(\frac{\partial E}{\partial \phi}\right)^*$ such that:
\be\label{adjoint3}
\left\langle\lam,\frac{\partial E}{\partial \phi}g\right\rangle=\left\langle\left(\frac{\partial E}{\partial \phi}\right)^*\lambda,g\right\rangle.
\ee
Hence, we define the directions $n$, $v$, $c$, and $s$ for the state variables $N$, $V$, $C$ and $S$. Let $g=[n, v, c, s]^T$, then
\[
\frac{\partial E}{\partial \phi}(\phi,p)g= \lim_{\mu\rightarrow0^+}\frac{E(\phi+\mu g,p)-E(\phi,p)}{\mu}.
\]
After some algebraics, it can be shown that $\frac{\partial E}{\partial \phi}\left(\phi,p\right)g$ is given by:
\be\label{constraintderiv}
\left[
\begin{array}{c}
n_t+\frac{V-yS^{\prime}}{S}n_y-\frac{{s}^{\prime} S-{S}^{\prime} s}{S^2}N_yy+N_y\frac{vS-Vs}{S^2}-[a-bN]n-N\left[\frac{\partial a}{\partial C}c-\frac{\partial b}{\partial C}Nc-bn\right]  
\\ [3mm]

v_y+\frac{2}{y}v-\frac{\partial b}{\partial C}NSc-bSn-bNs 
\\ [3mm]
c_{yy}+\frac{2}{y}c_y-kS^2n-\frac{\partial k}{\partial C}NS^2c-2kNSs \\ [3mm]
v(1,\cdot)-{s}^{\prime} \\ [1mm]
v(0,\cdot) \\ [1mm]
c(1,\cdot) \\ [1mm]
c_y(0,\cdot) \\ [1mm]
n(\cdot,0) \\ [1mm]
s(0)
\end{array}
\right].
\ee

Note that $E(\phi,p)$ and $\lambda$ should have the same number of components. Also, each component of $\lambda$ must be in a subspace of the corresponding component of $E$. For example, the first three components of $\lambda$ must depend on space and time, the fourth one only on time, the last one is just a real number, and so on.

So we define:
\begin{equation}\label{multiplier}
\lambda(y,t)=[\lam_1(y,t),\lam_2(y,t),\lam_3(y,t),\lam_4(t),\lam_5(t),\lam_6(t),\lam_7(t),\lam_8(y),\lam_9]^T.
\end{equation}

An inspection over equations (\ref{adjoint3}) and (\ref{constraintderiv}) shows that, roughly speaking, we should remove the spatial and temporal derivatives from $g$ and \textit{pass} them to $\lambda$.

The calculations make use of successive integration by parts to express each derivative of $g$ in terms of a derivative of $\lam$. Omitting here the details, that are shown in the Appendix, we obtain the following system of equations, which constitutes the adjoint problem (\ref{adjoint}):

\begin{eqnarray}
&\displaystyle-{\lam_1}_t-\left(\frac{V-y{S}^{\prime}}{S}\right){\lam_1}_y-\left(\frac{V_y-{S}^{\prime}}{S}+a-2bN\right)\lam_1-b S\lam_2-kS^2\lam_3=\mu_1(N^*-N),\label{adj1}\\
&\displaystyle{\lam_2}_y-\frac{2}{y}\lam_2-\frac{N_y}{S}\lam_1=0,\label{adj2}\\
&\displaystyle{\lam_3}_{yy}-\frac{2}{y}{\lam_3}_y+\left(\frac{2}{y^2}-\frac{\partial k}{\partial C}NS^2\right)\lam_3-\frac{\partial b}{\partial C}NS\lam_2-N\left(\frac{\partial a}{\partial C}-\frac{\partial b}{\partial C}N\right)\lam_1=0,\label{adj3}\\
&\lam_1(y,T)=0,\label{adj4}\\
&\lam_2(1,t)=-\lam_4(t),\label{adj6}\\
&{\lam_3}_y(0,t)=0,\label{adj7}\\
&\lam_3(1,t)=0,\label{adj8}\\
&\displaystyle {\lam_4}_t(t)=\int_0^1\left(\frac{N_{y}V}{S^2}\lam_1-\frac{y}{S}\frac{\partial}{\partial t}(N_y\lam_1)+bN\lam_2+2 kNS\lam_3\right)dy + \mu_2(S^*(t)-S(t)),\label{adj9}\\
&\displaystyle \lam_4(T)=-\int_0^1\frac{N_y(y,T)\lam_1(y,T)}{S(T)}y dy,\label{adj10}\\
& \lam_5(t)=\lam_2(0,t),\label{adj11}\\
& \lam_6(t)={\lam_3}_y(1,t)-2\lam_3(1,t),\label{adj12}\\
& \lam_7(t)=3\lam_3(0,t),\label{adj13}\\
& \lam_8(y)=\lam_1(y,0),\label{adj14}\\
&\displaystyle \lam_9=-\int_0^1\frac{ y N_y(y,0)\lam_1(y,0)}{S(0)} dy-\lam_4(0).\label{adj15}
\end{eqnarray}

Equations (\ref{adj1})-(\ref{adj15}) shall be solved in order to get $\lambda$.
Notice that the adjoint equations are posed backwards in time, with a terminal condition at $t=T$, while the state equations are posed forward in time, with an initial condition at $t=0$.

It is worth stressing that an explicit expression for $\lam_4$ can be obtained. In fact, taking into account that  $\lam_1(y,T)=0$, we have that
\[
 \int_t^T \frac{y}{S}\frac{\partial (N_y\lambda_1)}{\partial\tau} d\tau = \int_t^T N_y\frac{S'}{S^2}y\lam_1 d\tau - y\frac{N_y(y,t)\lam_1(y,t)}{S(t)}.
\]
Then, by equation \ref{eqN} from the direct problem, we have that

\[
 N_y\frac{S'}{S}y-\frac{V}{S}N_y=N_t-N[a-bN].
\]
But using the fact that $\lam_4(T)=0$, then $-\lam_4(t)=\int_t^T{\lam_4}_t d\tau$, i.e.,
\begin{eqnarray*}
\hspace*{-1cm}\lam_4(t)&=&\int_t^T\int_0^1\left(\frac{y}{S}\frac{\partial}{\partial t}(N_y\lam_1)-\frac{N_{y}V}{S^2}\lam_1-bN\lam_2-2 kNS\lam_3\right)dyd\tau + \mu_2\int_t^T(S-S^*)d\tau\\
&=&\int_t^T\int_0^1\left[\left(N_y\frac{S'}{S}y-\frac{V}{S}N_y\right)\frac{\lam_1}{S}-bN\lam_2-2 kNS\lam_3\right]dyd\tau - \int_0^1y\frac{N_y(y,t)\lam_1(y,t)}{S(t)}dy+ \mu_2\int_t^T(S-S^*)d\tau\\
&=&\int_t^T\int_0^1\left\{\left[N_t-N(a-bN)\right]\frac{\lam_1}{S}-bN\lam_2-2 kNS\lam_3\right\}dyd\tau - \int_0^1y\frac{N_y(y,t)\lam_1(y,t)}{S(t)}dy+ \mu_2\int_t^T(S-S^*)d\tau.
\end{eqnarray*}

\section{Designing an algorithm to solve the adjoint problem.}
\begin{center}
\begin{figure}
\includegraphics[width=16cm]{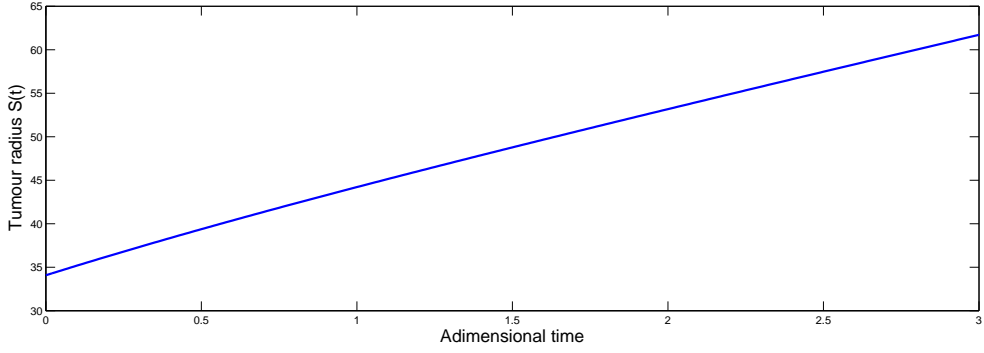}
\caption{Evolution of the tumour radius in time.}
\label{fig:directS_1}
\end{figure}
\end{center}

\begin{center}
\begin{figure}
\includegraphics[width=16cm]{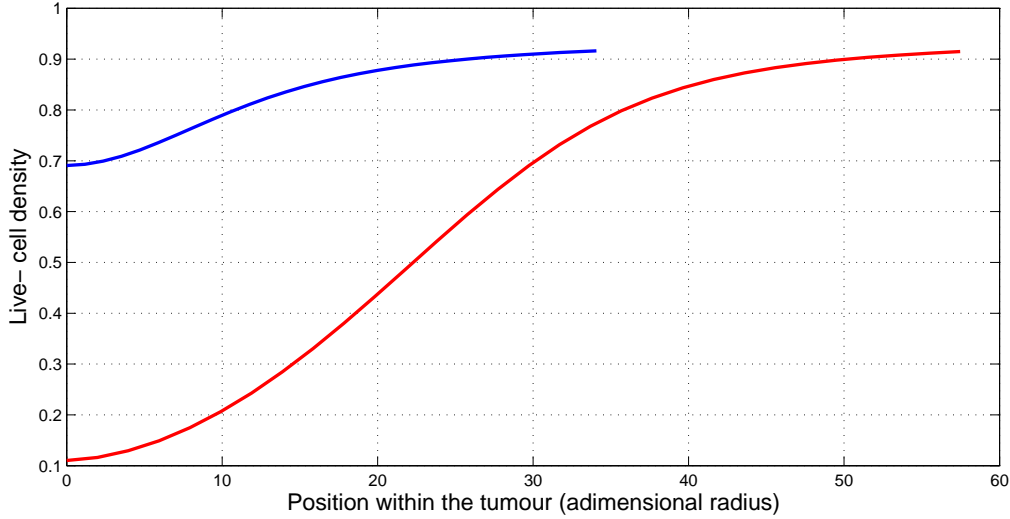}
\caption{Live-cell density within the tumour for two different times. Blue line corresponds to $t=0$ and red line corresponds to $t=2.5$. Note that from $t=0$ to $t=2.5$ the tumour has also grown in size.}
\label{fig:directN}
\end{figure}
\end{center}

It is worth stressing that obtaining model parameters via minimization of the objective functional $\tilde{J}$ is in general an iterative process requiring the value of the derivative. To compute $\tilde{J}'$ we just solve two systems of PDEs per iteration: the direct and the adjoint problems. This method is much cheaper than the sensitivity approach \cite{hinze} in which the direct problem is solved many times per iteration. That is why we developed an implementation in Fortran 2003 using an object-oriented strategy (with Fortran Intel Compiler 12.0.3). For the direct problem, Figure \ref{fig:directS_1} shows the evolution of the tumour radius in time, and Figure \ref{fig:directN} represents the living cell density within the tumour for two different times. 

Although the adjoint problem is quite similar to the direct one, there are more difficulties to solve it. For example, there is no explicit boundary condition for $\lam_2$. In our particular case, it is not necessary to compute $\left\{ \lam_i \right\}_{i=4}^9$ in order to calculate the derivative of $\tilde{J}$ with respect to $p$, because $\left\{ E_i \right\}_{i=4}^9$ does not depend on the parameters $p$ (see equations (\ref{constraint}) and (\ref{adjoint2})). However, $\lam_4$ is required since it gives us the boundary condition for $\lam_2$ (see equation (\ref{adj6})). To design a numerical procedure we perform the following steps at time $T$:
\begin{itemize}
\item[-] Equation (\ref{adj4}) states that $\lam_1(\cdot,T)=0$.
\item[-] By equation (\ref{adj10}) we have that $\lam_4(T)=0$, which gives us a boundary condition for $\lam_2$ (see equation (\ref{adj6})).
\item[-] Equation (\ref{adj2}) can be solved analytically getting $\lam_2(\cdot,T)=0$.
\item[-] Equations (\ref{adj3}), (\ref{adj7}) and (\ref{adj8}) allows us to obtain $\lam_3(\cdot,T)$.
\end{itemize}

Knowing the solution at time $t$, we obtain the solution at time $t-\Delta t$ in the following way:
\begin{itemize}
\item[-] By equation (\ref{adj1}) we first obtain ${\lam_1}_t(\cdot,t)$. Then we get $\lam_1(\cdot,t-\Delta t)$ using a backward finite difference.
\item[-] Using equation (\ref{adj9}), we integrate numerically to obtain ${\lam_4}_t(t)$ and then we get $\lam_4(t-\Delta t)$ by means of backward finite differences.
\item[-] With the value of $\lam_4(t-\Delta t)$ we obtain $\lam_2(1,t-\Delta t)$ via equation (\ref{adj6}).
\item[-] Equation (\ref{adj2}) can be solved numerically to get $\lam_2(\cdot,t-\Delta t)$.
\item[-] Solving equations (\ref{adj3}), (\ref{adj7}) and (\ref{adj8}) we obtain $\lam_3(\cdot,t-\Delta t)$.
\end{itemize}

As well as in the direct problem, in the adjoint one we have to be careful with the singularities in the PDEs. For example, if we take a look to equation (\ref{adj3}) together with the boundary conditions (\ref{adj7}) and (\ref{adj8}), for a fixed time $t$, we can ask ourselves about the solvability of this problem around $y=0$. However, there is a difference between the direct and the adjoint problems regarding to the  kind of singularities that equations (\ref{eqC}) and (\ref{adj3}) have in the origin.

The second term in (\ref{adj3}), for instance, looks harmless because ${\lam_3}_y(0,t)=0$ by (\ref{adj7}), so upon expanding ${\lam_3}_y$  by Taylor about $0$ and dividing by $y$, the singularity disappears. On the other hand, the problem with the third term is harder, because we get a blowup in the origin. To solve this problem we transform equations (\ref{adj3}), (\ref{adj7}) and (\ref{adj8}) into a first order ODE for a fixed time $t$, namely:
\be\label{sistema}
\left[\begin{array}{c}
u \\
v \end{array}\right]^{\prime} = \left[\begin{array}{c} v \\ [3mm]
\displaystyle \frac{2}{y}v-\left(\frac{2}{y^2}-NS^2\frac{\partial k}{\partial C}\right)u-\left(N^2\frac{\partial b}{\partial C}-N\frac{\partial a}{\partial C}\right)\lam_1+NS\frac{\partial b}{\partial C}\lam_2 \end{array}\right],
\ee
\be\label{sistemabc1}
u(1)=0,
\ee
\be\label{sistemabc2}
v(0)=0,
\ee
where $u(y)=\lam_3(y,t)$ and $v(y)={\lam_3}_y(y,t)$. Then, for a fixed $\epsilon>0$ we propose a parameter $q=v(1)$, and solve the system (\ref{sistema})-(\ref{sistemabc2}) in the interval $[\epsilon,1]$ with boundary conditions $u(1)=0$ and $v(1)=q$, obtaining a solution $[u_q, v_q]^T$. Using Taylor expansions near $y=0$ we extend these solutions to the whole interval $[0,1]$ (see \cite{ascher}).
\begin{eqnarray}
u_q(0)&\approx& u_q(\epsilon)-\epsilon u'_q(\epsilon), \label{taylor1}
\\
v_q(0)&\approx& v_q(\epsilon)-\epsilon v'_q(\epsilon). \label{taylor2}
\end{eqnarray}

The next step is to define a function
\be\label{bisection function}
F(q)= v_q(0),
\ee
and to find a root of $F$, i.e. to find $\hat{q}$ such that $F(\hat{q})=0$. Then, the solution of the system will be $[u_{\hat{q}}, v_{\hat{q}}]^T$ extended to the interval $[0,1]$.

To solve the first order ODE (\ref{adj2}) for $\lam_2(\cdot,t)$ with boundary condition $\lam_2(1,t)$ known from (\ref{adj6}), we also solve the problem in the interval $[\epsilon,1]$ and then extend the solution to the interval $[0,1]$ using a first order Taylor expansion.

In general, the derivatives that appear in the adjoint system of PDEs are approximated using a finite difference scheme. For example, in order to solve equation (\ref{adj1}) we consider
\[
\lambda_1(y,t-\Delta t)\approx\lam_1(y,t)-{\lam_1}_t(y,t)\Delta t,
\]
and using (\ref{adj1}) we get
\begin{eqnarray*}
\lam_1(y,t-\Delta t)&\approx&\lam_1(y,t)+ \Delta t \left(\frac{{S^{\prime}(t)}}{S(t)}y+\frac{V(y,t)}{S(t)}\right){\lam_1}_y
\\
&+&\Delta t \left(\frac{{S^{\prime}(t)}}{S(t)}+\frac{V_y(y,t)}{S(t)}+a(C(y,t))-2b(C(y,t))N(y,t)\right)\lam_1(y,t) 
\\
&+&\Delta t \left\lbrace b(C(y,t))S(t)\lam_2(y,t)+k(C(y,t))S(t)^2\lam_3(y,t)+N^*(y,t)-N(y,t)\right\rbrace.
\end{eqnarray*}

\section{Optimization.}

It is well-known \cite{nocedal} that gradient-based optimization algorithms require the evaluation of the gradient of the functional. One important advantage of evaluating the gradient through adjoints is that it requires to solve the adjoint problem only once per iteration, regardless the number of inversion variables. Note that the derivative of the functional can be approximated by using finite differences, but this is an expensive approach because it needs, for each optimization iteration, to solve the forward problem as many times as inversion variables are.

The method we will use for minimizing the functional $\tilde{J}$ can be summarized as follows:

\begin{algorithm}\label{alg}
\textbf{Adjoint-based minimization method.}

\begin{enumerate}
\item[1.] Give an initial guess $p^0$ for the vector of parameters.
\item[2.] Given the vector $p^k$ in step $k$, solve the direct and adjoint problems at this step.
\item[3.] Obtain the derivative of the functional, i.e. $\tilde J'(p^k)$, using (\ref{adjoint2}).
\item[4.] Move in the direction of $-\tilde J'(p^k)$, i.e., compute $p^{k+1}=\Pi_{U_{ad}}\left[p^{k}-\alpha\tilde J ' (p^k)\right]$, where $\alpha$ is a positive parameter to be chosen, and $\Pi_{U_{ad}}$ denotes the projection on the set of admissible points.
\item[5.] Stop when $\tilde{J}\left( p^{k+1} \right)$ is less than a tolerance $TOL_1>0$, or when the distance between two consecutive iterations is less than a tolerance $TOL_2>0$, that is, $\parallel p^{k+1}-p^{k} \parallel < TOL_2$
\end{enumerate}
\end{algorithm}

\section{Numerical experiments.}
The goal of this section is to test and evaluate the performance of an adjoint-based optimization method, by executing some numerical simulations of Algorithm \ref{alg} for some test-cases.

The living cell density and the tumour radius are generated via the forward problem. We show here the results obtained by assuming \textit{standard values} $c_c=0.1$, $c_d=0.05$, $\sigma=0.9$, as suggested in \cite{WK1}. 

\begin{center}
\begin{figure}
\includegraphics[height=4.3cm]{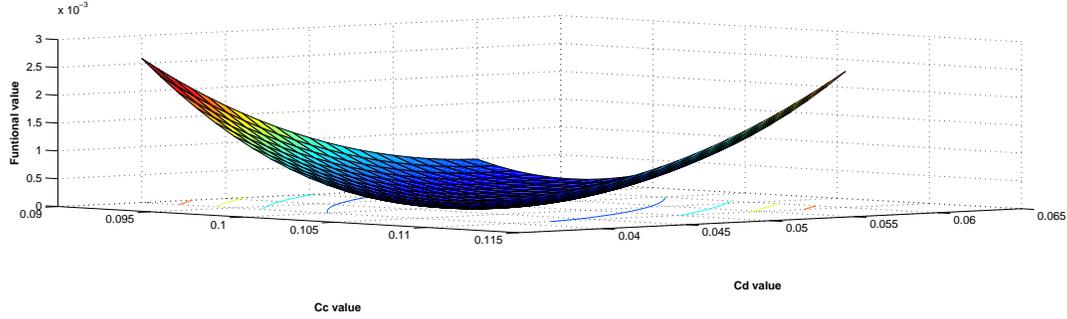}
\caption{Functional value $J$ in terms of $c_c$ and $c_d$ for constant $\sigma=0.9$. Note that the surface reaches a minimum near $c_c=0.1$ and $c_d=0.05$}
\label{fig:figureadj5}
\end{figure}
\end{center}

\subsection{Model-generated data.}

Consider first an optimization problem that consist in minimizing
the functional (\ref{jota2}), where $N^*(y,t)$ and $S^*(t)$ are
generated via the forward model, for a choice of the model parameters $c_c=0.1$, $c_d=0.05$, $\sigma=0.9$.

The tumour is first detected at time $t=0$, by which time it has grown following the model \cite{WK1}. Originally, at an adimensional time $t\approx -4.5$, a single cell started to take nutrients from the environment, letting it grow up to a dimensionless size $S_I\approx 34$. Thus, the initial profile for this test case is the one shown in figures \ref{fig:directS_1} and \ref{fig:directN}.  

In order to work with functional (\ref{jota2}), it is necessary to define the landmark points $y_j$ and the times $t_k$ where the measurements are made. For simplicity, and to be consistent with the way we solved the direct problem, we took the same spatial grid for the landmarks, i.e., 30 equidistant points $0=y_1<...<y_{30}=1$. Regarding to the time selection, it is apparent from the experiments that taking the adimensional time $T=0.5$ is sufficiently representative (it corresponds to 50 time steps of length 0.01). The factors $\mu_1$ and $\mu_2$ are taken to be $100$ and $1$ respectively, and the parameter $\alpha$ used in the projection over the admissible set is taken to be $0.1$.

Figure \ref{fig:figureadj5} shows the value that the functional (\ref{jota2}) takes for different values of $c_c$ and $c_d$, remaining $\sigma$ as a constant. It is worth mentioning that $J$ looks convex and that the variations are greater with respect to $c_c$ compared to those with respect to $c_d$.

The idea of this test case is to investigate how close the original
value of the parameter can be retrieved. However, it is not a
trivial one, because we do not know, for instance, if the
optimization problem has a solution or, in that case, if it is
unique or if the method converges to another local minima.

We emphasize that we have run the algorithm several times using different
initial random conditions and in all cases the results were similar.
They can be summarized as:

\begin{itemize}

\item[-] Stopping criteria: \textit{functional value lower than $10^{-6}$ or norm of the gradient lower than $10^{-12}$}
\item[-] Iterations/elapsed time: \textit{140/35 min}
\item[-] Initial point: $p_0=[0.16, 0.03, 1.0]$
\item[-] Final point: $p_f=[0.1006492, 0.084465653, 0.9297853]$
\item[-] Functional final value: \textit{$J(p_f)=0.991496220 \times 10^{-6}$}

\end{itemize}

\begin{center}
\begin{figure}
\includegraphics[width=16cm]{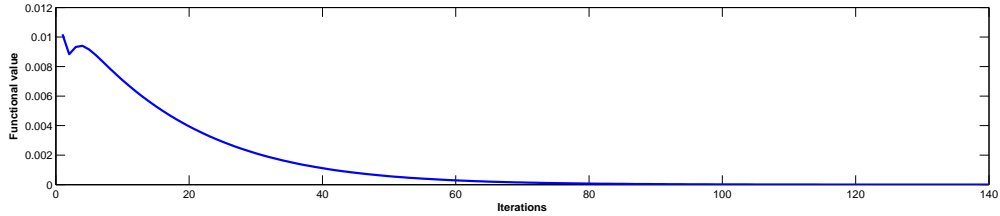}
\caption{Evolution of the functional value with the number of iterations.}
\label{fig:figureadj1}
\end{figure}
\end{center}

\begin{center}
\begin{figure}
\includegraphics[width=16cm]{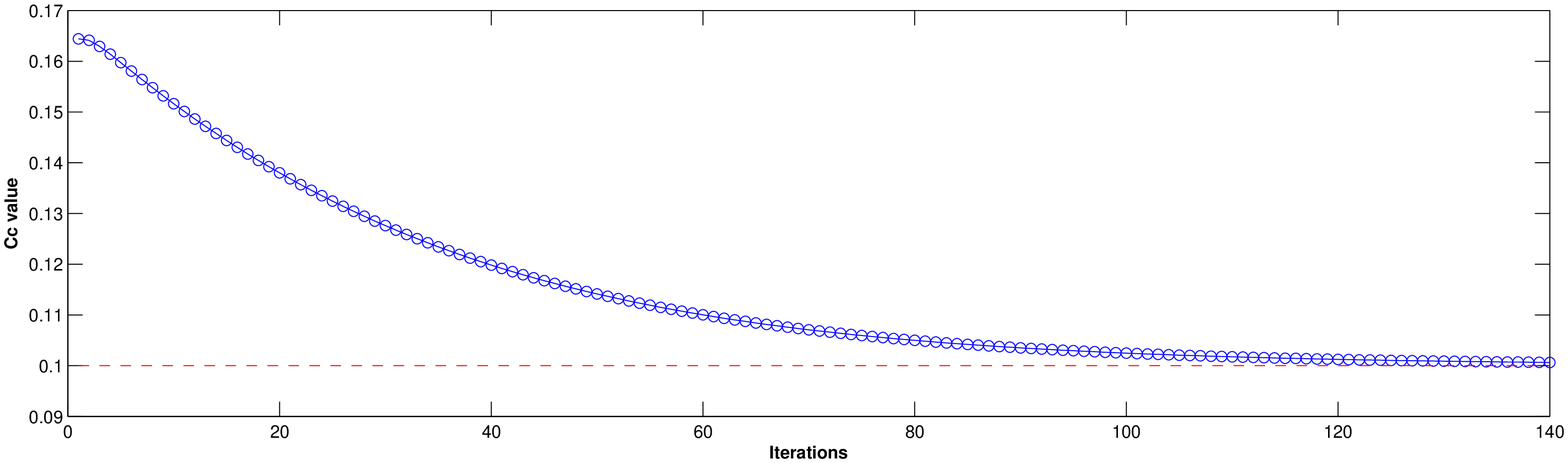}
\caption{Evolution of the $c_c$ value with the number of iterations (blue line) and \textit{real value} of $c_c$ (red line).}
\label{fig:figureadj2}
\end{figure}
\end{center}

\begin{center}
\begin{figure}
\includegraphics[width=16cm]{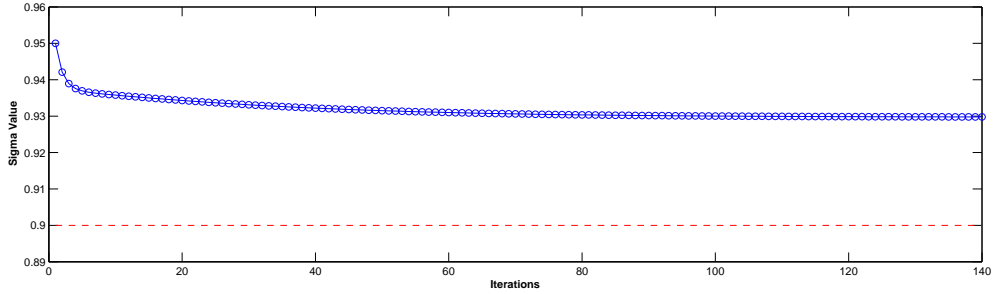}
\caption{Evolution of the $\sigma$ value with the number of iterations (blue line) and \textit{real value} of $\sigma$ (red line).}
\label{fig:figureadj3}
\end{figure}
\end{center}

Figure \ref{fig:figureadj1} represents the evolution in the value of $J$ with the number of iterations. It is worth stressing that, even the stopping criteria required $140$ iterations, taking just about $90$ iterations would be sufficient to obtain similar results. In fact, Figures \ref{fig:figureadj2} and \ref{fig:figureadj3} show the evolution of $c_c$ and $\sigma$ respectively, and the \textit{real value} of this parameters.

\subsection{Model-generated data with random noise.}

It is well known that the presence of noise in the data may imply
the appearance of strong numerical instabilities in the solution of
an inverse problem \cite{bertero-piana}.

The outputs of the detectors and experimental equipments where
the variables $N^*$ and $S^*$ are measured are often affected by
perturbations, usually random ones. As stated in \cite{AgBaTu}, it
is in general valid to consider a $5\%$ of random noise.

Once again, it is assumed that the tumour is first detected at time $t=0$, by which time it has grown following the model \cite{WK1}. Originally, at an adimensional time $t\approx -7$, a single cell started to take nutrients from the environment, letting it grow up to a dimensionless size $S_I\approx 53$.  

The landmark points, the adimensional time $T=0.5$ and the factors $\mu_1$, $\mu_2$ and $\alpha$ are taken to be equal as in the previous case. Note that, although this requires more iterations and consequently more computational time, we would get more information from a tumour that was detected a little bit later.

After running the algorithm several times using different
initial random conditions, the obtained results were similar.
They can be summarized as:

\begin{itemize}

\item[-] Stopping criteria: \textit{functional value lower than $10^{-6}$ or norm of the gradient lower than $10^{-12}$}
\item[-] Iterations/elapsed time: \textit{140/35 min}
\item[-] Initial point: $p_0=[0.08, 0.07, 0.93]$
\item[-] Final point: $p_f=[0.1105396, 0.07723431, 0.9172613]$
\item[-] Functional final value: \textit{$J(p_f)=0.01830000000$}

\end{itemize}

\begin{center}
\begin{figure}
\includegraphics[width=16cm]{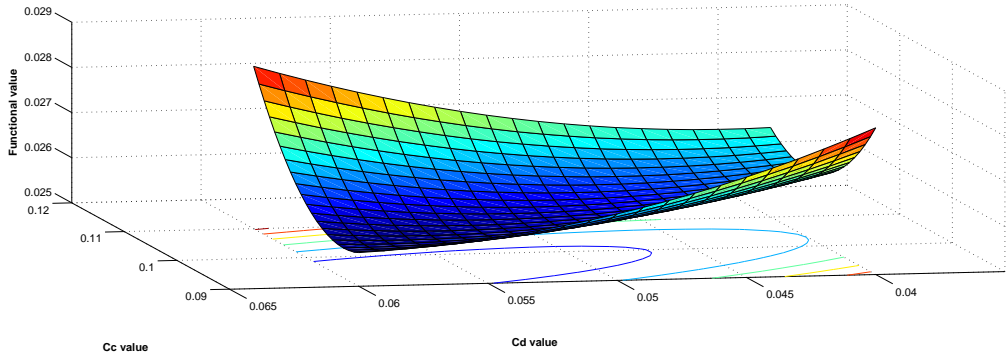}
\caption{Functional value $J$ in terms of $c_c$ and $c_d$ for constant $\sigma=0.9$, for data with a $5\%$ of random noise.}
\label{fig:figure7}
\end{figure}
\end{center}

Figure \ref{fig:figure7} shows the value that the functional (\ref{jota2}) takes for different values of $c_c$ and $c_d$, remaining $\sigma$ as a constant and assuming that $S^*$ and $N^*$ are obtained with a $5\%$ of random noise.

\begin{center}
\begin{figure}
\includegraphics[width=16cm]{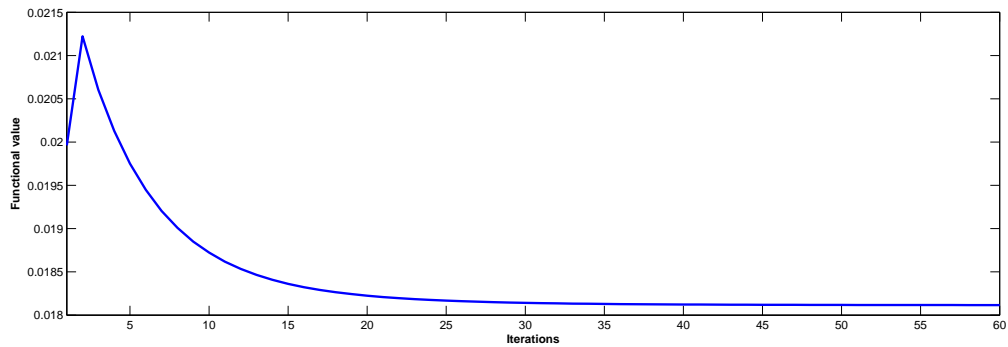}
\caption{Evolution of the functional value with the number of iterations, for data with $5\%$ of random noise.}
\label{fig:figure9}
\end{figure}
\end{center}

\begin{center}
\begin{figure}
\includegraphics[width=16cm]{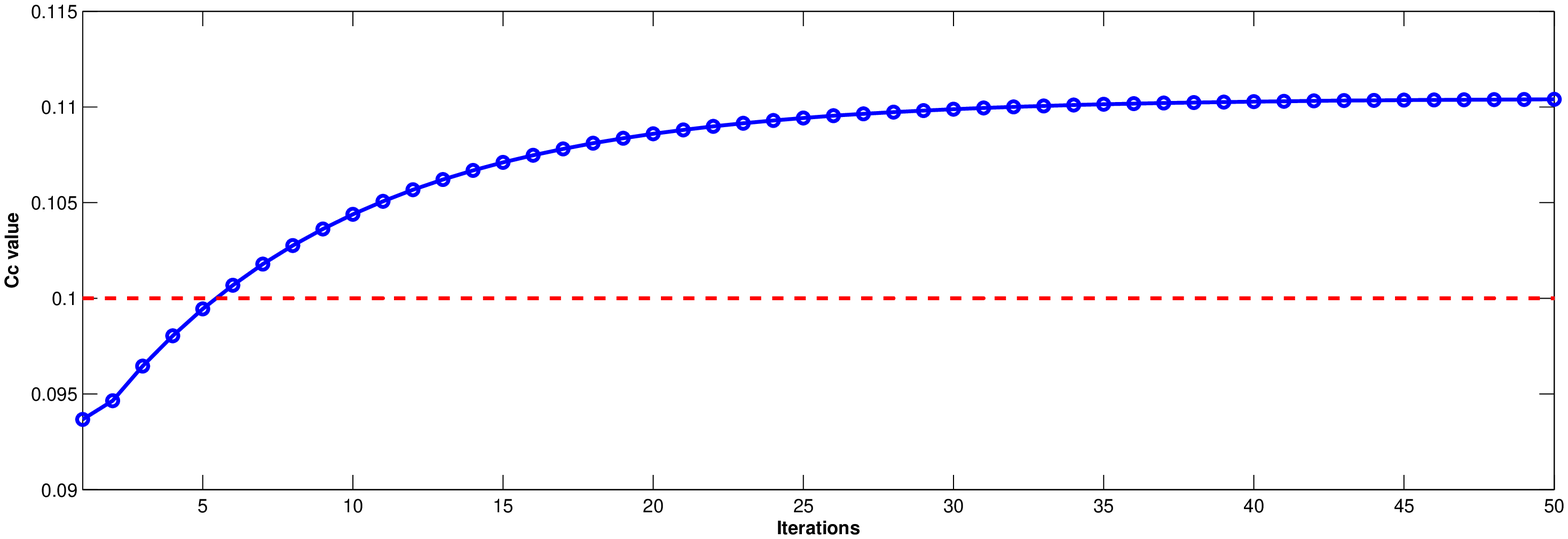}
\caption{Evolution of the $c_c$ value with the number of iterations (blue line) for data with $5\%$ of random noise and \textit{real value} of $c_c$ (red line).}
\label{fig:figure8bis}
\end{figure}
\end{center}

Figure \ref{fig:figure9} represents the evolution in the value of $J$ with the number of iterations. A comparison with Figure \ref{fig:figureadj1} shows that the functional values are greater in this case, but the algorithm stops because the variations become small. Figure \ref{fig:figure8bis} shows the evolution of $c_c$ and the \textit{real value} of this parameter.

\begin{center}
\begin{figure}
\includegraphics[width=16cm]{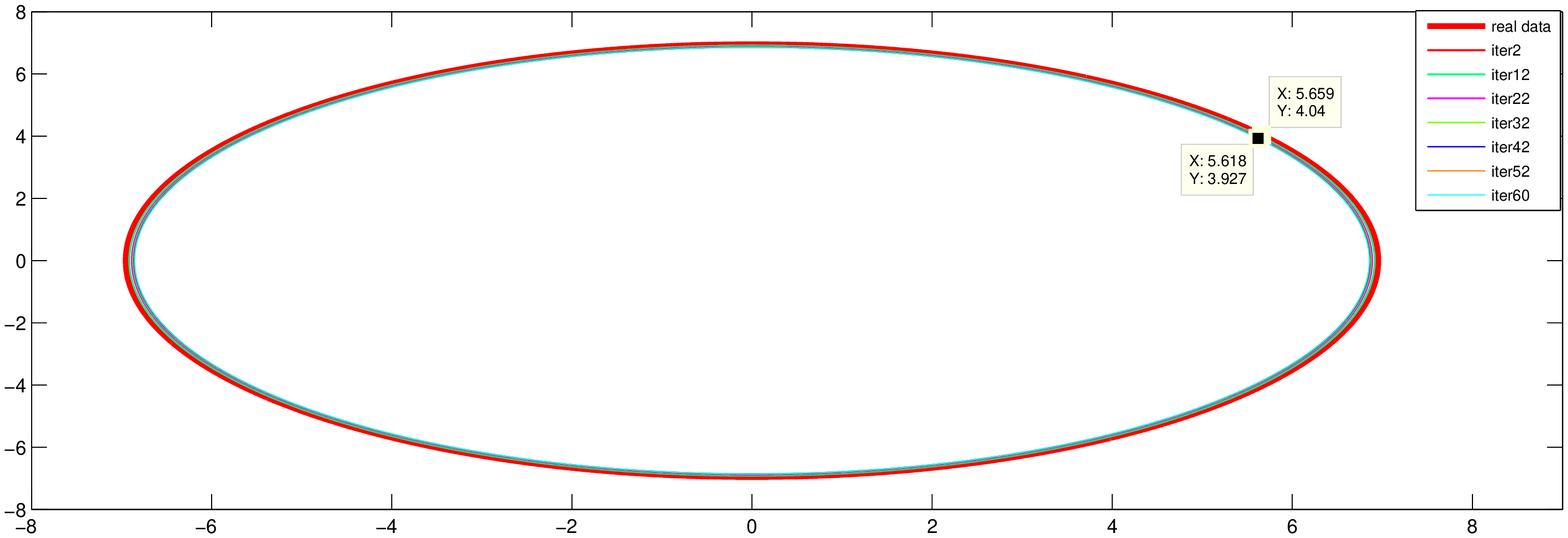}
\caption{A sequence of spheroids obtained by Algorithm \ref{alg} and choice of one point in the boundary of the real boundary and other point in the boundary of the tumour obtained after 60 iterations.}
\label{fig:figure10}
\end{figure}
\end{center}

\begin{center}
\begin{figure}
\includegraphics[scale=0.5]{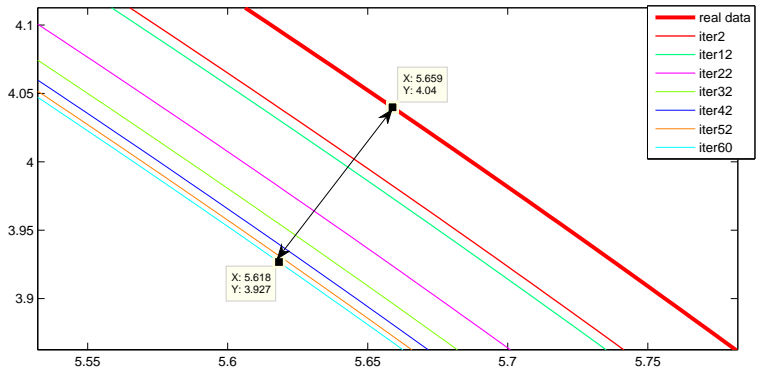}
\caption{A zoom from Figure \ref{fig:figure10} that shows the real tumour radius and the one obtained after 60 iterations.}
\label{fig:figure10zoom}
\end{figure}
\end{center}

We can choose one of the variables considered in the functional (\ref{jota2}) and look for difference between the real value of this variable and the value that corresponds to the solution of the direct problem for the parameters obtained after running the algorithm. For example, let us take the tumour radius at time T. Figure \ref{fig:figure10} shows a sequence of spheroids obtained in some of the iterations. As the difference between the radii is small compared to the radii themselves, we make a zoom, obtaining Figure \ref{fig:figure10zoom}.

\subsection{Comparison between the two cases with and without noise.}

The case in which we considered model-generated data with $5\%$ of random noise is, as expected, not as precise as the case in which the data is generated without noise.

\begin{figure}
\begin{center}
\includegraphics[scale=0.5]{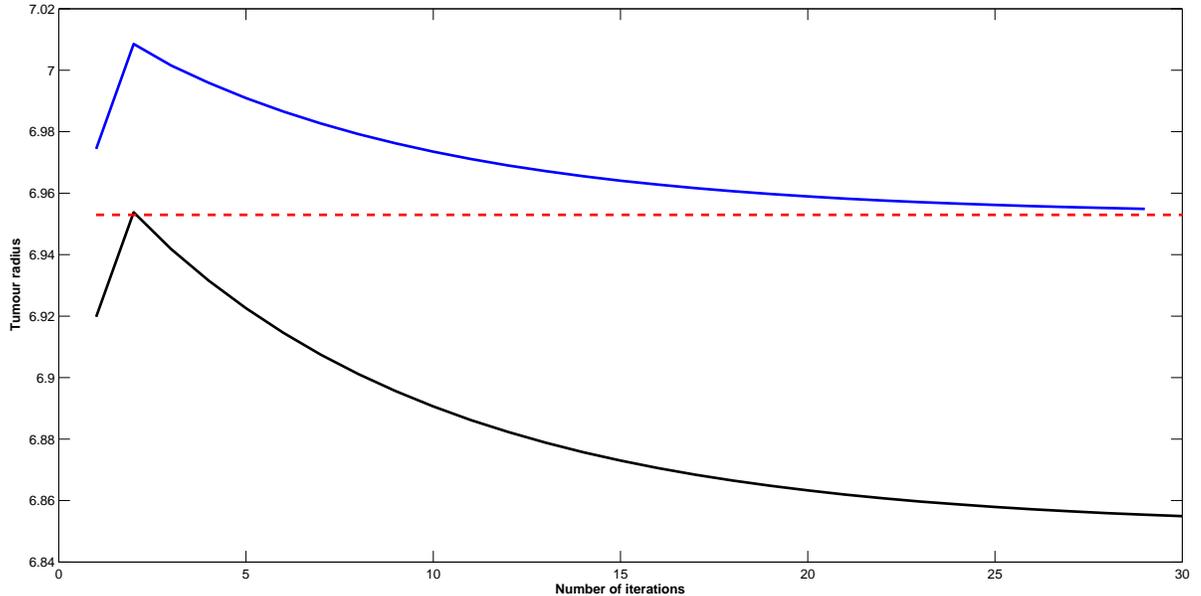}
\end{center}
\setlength{\belowcaptionskip}{2pt} \caption{Evolution of the tumour radius for each iteration: blue line corresponds to model-generated data without noise, and black line to model-generated data with a $5\%$ of random noise. Red line represents the real observed radius at time T.}
\label{fig:figure8}
\end{figure}

First of all, Algorithm \ref{alg} stops by a different reason: the functional can reach lower values in the case without noise than in the other one, so the derivative of the functional becomes an important stopping criteria. However, it is worth stressing that even with the presence of noise in the data, the method let us compute the parameters with a small error, in some cases of about $1\%$. Figure \ref{fig:figure8} shows the evolution of the tumour radius at time $T$ for each iteration, comparing both cases with the \textit{real radius}.

\section{Conclusions.}

The scientific community agrees that life's sciences, like biology or medicine, need the development of new tools in order to build models able to reproduce and to predict real phenomena. Over the last decades, a number of mathematical models for cancer onset and growth have been proposed \cite{Adambellomo, BellomoLiMaini}, and it became clear that these models are expected to success if the parameters involved in the modeling process are known. Or eventually, taking into account that some biological parameters may be unknown (especially in vivo), the model can be used to obtain them \cite{AgBaTu, DoAsPa11}. 

This paper, as already mentioned in Section 1, aims at offering a mathematical tool for the obtention of phenomenological parameters which can be identified by inverse estimation, by making suitable comparisons with experimental data. The inverse problem was stated as a PDE-constrained optimization problem, which was solved by using the adjoint method. The adjoint-based technique - although mathematically more complicated that the pattern search method used in \cite{knopoff} - has shown to work more efficiently, obtaining the results with better accuracy and with a less expensive numerical resolution. In addition, the gradient of the proposed functional is obtained and can be extended, in principle, to any number of unknown parameters.  

Focusing on further developments of the mathematical tools, it is worth mentioning that the numerical resolution proposed in this paper is in some aspects challenging and several numerical procedures were introduced in order to deal with non-linearities and singularities in the adjoint system of PDEs.

In addition, we remark that the parameter estimation via PDE-constrained optimization is a general approach that can be used, for instance, to consider the effects of chemotherapy. We are currently working in this line and also in the resolution of the optimization problem but after discretizing the original system of PDEs, in order to compare and contrast the performance of both methods.

\section*{Acknowledgments.}
We appreciate the courtesy of Milena Batalla and Lucia Policastro, from the \textit{Grupo de Micro y Nanotecnolog\'ia - Gerencia de \'Area de Investigaciones y Aplicaciones No Nucleares - CNEA Argentina}, who strongly contributed with information and motivations for this work. 

The work of  the authors  was partially supported by grants from
CONICET, SECYT-UNC and PICT-FONCYT.

% \bibliographystyle{abbrv}
% \bibliography{adjoint_referencias}

\begin{thebibliography}{10}

\bibitem{Adambellomo}
J.~Adam and N.~Bellomo.
\newblock {\em A survey of models for tumor immune systems dynamics}.
\newblock Modeling and simulation in science, engineering \& technology.
  Birkh{\"a}user, 1997.

\bibitem{Adam1}
J.~A. Adam.
\newblock A simplified mathematical model of tumor growth.
\newblock {\em Mathematical Biosciences}, 81(2):229 -- 244, 1986.

\bibitem{AgBaTu}
J.~Agnelli, A.~Barrea, and C.~Turner.
\newblock Tumor location and parameter estimation by thermography.
\newblock {\em Mathematical and Computer Modelling}, 53(7-8):1527--1534, 2011.

\bibitem{ascher}
U.~M. Ascher, R.~M.~M. Mattheij, and R.~D. Russell.
\newblock {\em Numerical solution of boundary value problems for ordinary
  differential equations}, volume~13 of {\em Classics in Applied Mathematics}.
\newblock Society for Industrial and Applied Mathematics (SIAM), Philadelphia,
  PA, 1995.
\newblock Corrected reprint of the 1988 original.

\bibitem{BeChaDe09}
N.~Bellomo, M.~Chaplain, and E.~De~Angelis.
\newblock {\em Selected Topics on Cancer Modeling - Genesis - Evolution -
  Immune Competition - Therapy}.
\newblock Birkh{\"a}user, Boston, 2009.

\bibitem{BellomoLiMaini}
N.~Bellomo, N.~Li, and P.~Maini.
\newblock On the foundations of cancer modelling: selected topics,
  speculations, and perspectives.
\newblock {\em Mathematical Models and Methods in Applied Sciences},
  18(04):593--646, 2008.

\bibitem{bertero-piana}
M.~Bertero and M.~Piana.
\newblock {Inverse problems in biomedical imaging: modeling and methods of
  solution}.
\newblock {\em Complex Systems in Biomedicine}, pages 1--33, 2006.

\bibitem{branden}
C.~Brandenburg, F.~Lindemann, M.~Ulbrich, and S.~Ulbrich.
\newblock A continuous adjoint approach to shape optimization for {N}avier
  {S}tokes flow.
\newblock In {\em Optimal control of coupled systems of partial differential
  equations}, volume 158 of {\em Internat. Ser. Numer. Math.}, pages 35--56.
  Birkh\"auser Verlag, Basel, 2009.

\bibitem{Byrne2}
H.~Byrne and M.~Chaplain.
\newblock {Free boundary value problems associated with the growth and
  development of multicellular spheroids}.
\newblock {\em European Journal of Applied Mathematics}, 8(06):639--658, 1997.

\bibitem{Crank}
J.~Crank.
\newblock {\em Free and moving boundary problems}.
\newblock Oxford Science Publications. The Clarendon Press Oxford University
  Press, New York, 1984.

\bibitem{engl}
H.~W. Engl, M.~Hanke, and A.~Neubauer.
\newblock {\em Regularization of inverse problems}, volume 375 of {\em
  Mathematics and its Applications}.
\newblock Kluwer Academic Publishers Group, Dordrecht, 1996.

\bibitem{Greenspan}
H.~Greenspan.
\newblock {Models for the growth of a solid tumor by diffusion}.
\newblock {\em Stud. Appl. Math}, 51(4):317--340, 1972.

\bibitem{hinze}
M.~Hinze, R.~Pinnau, M.~Ulbrich, and S.~Ulbrich.
\newblock {\em Optimization with {PDE} constraints}, volume~23 of {\em
  Mathematical Modelling: Theory and Applications}.
\newblock Springer, New York, 2009.

\bibitem{hogea}
C.~Hogea, C.~Davatzikos, and G.~Biros.
\newblock An image-driven parameter estimation problem for a reaction-diffusion
  glioma growth model with mass effects.
\newblock {\em J. Math. Biol.}, 56(6):793--825, 2008.

\bibitem{kirsch}
A.~Kirsch.
\newblock {\em An introduction to the mathematical theory of inverse problems},
  volume 120 of {\em Applied Mathematical Sciences}.
\newblock Springer-Verlag, New York, 1996.

\bibitem{knopoff}
D.~A. Knopoff, D.~R. Fern\'andez, G.~A. Torres, and C.~V. Turner.
\newblock {A parameter estimation problem for a tumour growth model }.
\newblock submitted (2011).

\bibitem{nocedal}
J.~Nocedal and S.~J. Wright.
\newblock {\em Numerical optimization}.
\newblock Springer Series in Operations Research. Springer-Verlag, New York,
  1999.

\bibitem{PeZu07}
B.~Perthame and J.~P. Zubelli.
\newblock On the inverse problem for a size-structured population model.
\newblock {\em Inverse Problems}, 23(3):1037--1052, 2007.

\bibitem{cnea}
G.~A. Santa~Cruz, S.~J. Gonz\'alez, A.~Dagrosa, A.~E. Schwint, M.~Carpano,
  V.~A. Trivillin, E.~F. Boggio, J.~Bertotti, J.~Mar\'in, A.~Monti~Hughes,
  A.~J. Molinari, and M.~Albero.
\newblock Dynamic infrared imaging for biological and medical applications in
  boron neutron capture therapy.
\newblock {\em Thermosense: Thermal Infrared Applications XXXIII}, 8013, 2011.

\bibitem{Shymko}
R.~Shymko and L.~Glass.
\newblock {Cellular and geometric control of tissue growth and mitotic
  instability* 1}.
\newblock {\em Journal of Theoretical biology}, 63(2):355--374, 1976.

\bibitem{DoAsPa11}
K.~van~den Doel, U.~M. Ascher, and D.~K. Pai.
\newblock Source localization in electromyography using the inverse potential
  problem.
\newblock {\em Inverse Problems}, 27(2):025008, 20, 2011.

\bibitem{WK1}
J.~P. Ward and J.~R. King.
\newblock Mathematical modelling of avascular-tumour growth.
\newblock {\em Mathematical Medicine and Biology}, 14(1):39--69, 1997.

\bibitem{WK2}
J.~P. Ward and J.~R. King.
\newblock Mathematical modelling of drug transport in tumour multicell
  spheroids and monolayer cultures.
\newblock {\em Mathematical Biosciences}, 181(2):177 -- 207, 2003.

\bibitem{ZuMa03}
J.~P. Zubelli, R.~Marabini, C.~O.~S. Sorzano, and G.~T. Herman.
\newblock Three-dimensional reconstruction by chahine's method from electron
  microscopic projections corrupted by instrumental aberrations.
\newblock {\em Inverse Problems}, 19(4):933--949, 2003.

\end{thebibliography}

\newpage

\appendix

\section*{Appendix\\ Obtaining the adjoint problem.}

In this section we show the calculations involved in order to obtain the adjoint equations (\ref{adj1})-(\ref{adj15}).
As stated in Section \ref{sec:adjoint}, the adjoint equations constitute a system of PDEs, with unknown $\lam$, given by (\ref{adjoint}). To ease calculations, let us fix $g=[n,v,c,s]^T\in\mathcal{Y}$. Hence, by (\ref{jota2}), we have that
\be\label{jota2deriv}
\frac{\partial J}{\partial \phi}g=\mu_1\int_0^1\int_{0}^T [N
(y,t)-N^*(y,t)] n(y,t) dt dy + \mu_2 \int_{0}^T [S(t)-S^*(t)] s(t) dt.
\ee

On the other hand, $(\frac{\partial E}{\partial \phi})^*\lam$ is obtained by using (\ref{adjoint3}). In what follows, we shall obtain equivalent expressions for each of the nine terms of the summation $\langle \frac{\partial E}{\partial \phi}g,\lam\rangle$, which are associated with the nine constraints given by $E$ in (\ref{constraint}).

\paragraph{Constraint 1}

Consider the expression
\[
\begin{array}{r}
\displaystyle\int_0^1\int_{0}^T\left[n_t+\frac{V-yS^{\prime}}{S}n_y -\frac{{s}^{\prime} S-{S}^{\prime} s}{S^2}N_y y\right.+N_y\frac{vS-Vs}{S^2} -\left(a-bN\right)n\qquad\\
\displaystyle-\left.N\left(\frac{\partial a}{\partial C}c-\frac{\partial b}{\partial C}Nc-bn\right)\right]\lam_1 dt dy.
\end{array}
\]

Using integration by parts repeatedly and the facts that $V(1,t)=S^\prime (t)$ and $V(0,t)=0$, it yields
{\setlength\arraycolsep{0.1em}
\begin{eqnarray}
\int_0^1\int_{0}^T \left[-{\lam_1}_t+\frac{{S}^{\prime}}{S}\lam_1\right.& +&\left.y\frac{{S}^{\prime}}{S}{\lam_1}_y -\frac{V_y}{S}\lam_1-\frac{V}{S}{\lam_1}_y-\left(a-2bN\right)\lam_1\right]n\,dt dy \nonumber \\
&+&\int_0^1\int_{0}^T \frac{N_y}{S}\lam_1 v\, dt dy \nonumber \\
&-&\int_0^1\int_{0}^T N\left(\frac{\partial a}{\partial C}-\frac{\partial b}{\partial C}N\right)\lam_1 c\, dt dy \nonumber \\
&+&\int_0^1 \int_{0}^T \left(\frac{y N_{yt}}{S}\lam_1 + \frac{y N_y}{S}{\lam_1}_t-\frac{N_y V}{S^2}\lam_1\right) s\, dt dy \label{constraint1} \\
&+&\int_0^1\left(\lam_1(y,T)n(y,T)-\lam_1(y,0)n(y,0)\right)dy \nonumber\\
&+&s(0) \int_0^1\frac{y N_y(y,0)\lam_1(y,0)}{S(0)}dy  \nonumber \\
&-&s(T) \int_0^1\frac{y N_y(y,T)\lam_1(y,T)}{S(T)}dy.\nonumber
\end{eqnarray}
}

\paragraph{Constraint 2}

We have the following expression

\be\nonumber
\int_0^1\int_{0}^T\left(v_y+\frac{2}{y}v-\frac{\partial b}{\partial C}NSc-bSn-bNs\right)\lam_2 dtdy.
\ee

In this case we have to integrate by parts just in the first term, because it is the only one that has a derivative of $g$, in this case $v_y$. So we obtain that the second term in the inner product is
\be\label{constraint2}
\begin{array}{r}
\displaystyle \int_0^1\int_{0}^T\left[\left(-{\lam_2}_y+\frac{2}{y}\lam_2\right)v-\frac{\partial b}{\partial C}NS\lam_2 c-bS\lam_2 n-bN\lam_2 s\right] dtdy\qquad\\
+\displaystyle \int_{0}^T \left(\lam_2(1,t)v(1,t)-\lam_2(0,t)v(0,t)\right)dt.
\end{array}
\ee

\paragraph{Constraint 3}

The expression to be taken into account is

$$
\int_0^1\int_{0}^T\left(c_{yy}+\frac{2}{y}c_y-kS^2n-\frac{\partial k}{\partial C}NS^2c-2kNSs\right)\lam_3 dtdy
$$

Because of the presence of second order derivatives we shall perform integration by parts twice. First of all, note that
\[
c_{yy}+\frac{2}{y}c_y= \frac{1}{y}\frac{\partial^2}{\partial y^2}(yc).
\]
Then, we have,
\begin{eqnarray*}
\int_0^1\left(c_{yy}+\frac{2}{y}c_y\right)\lam_3 dy&=&\int_0^1 \frac{\lam_3}{y}\frac{\partial^2}{\partial y^2}(yc) dy\\
&=&\left.\frac{\lam_3}{y}\frac{\partial}{\partial y}(yc)\right|_{y=0}^{y=1} - \left.yc\frac{\partial}{\partial y}\left(\frac{\lam_3}{y}\right)\right|_{y=0}^{y=1} + \int_0^1 yc\frac{\partial^2}{\partial y^2}\left(\frac{\lam_3}{y}\right)dy.
\end{eqnarray*}

To evaluate the limits, we assume that $\lam_3(y,t)c(y,t)\rightarrow 0$ when $y\rightarrow 0$ and apply l'H\^opital's rule. Hence, the third term is equal to
{\setlength\arraycolsep{0.1em}
\begin{eqnarray}
\int_0^1\int_{0}^T \left({\lam_3}_{yy}-\frac{2{\lam_3}_y}{y}\right.&+&\left.\frac{2\lam_3}{y^2}\right)c-\left(kS^2n+\frac{\partial k}{\partial C}NS^2c+2kNSs\right)\lam_3\, dt dy\nonumber\\
&+&\int_{0}^T\left(2\lam_3(1,t)c(1,t)+\lam_3(1,t)c_y(1,t)-{\lam_3}_y(1,t)c(1,t)\right)dt\label{constraint3}\\
&-&\int_{0}^T \left(3\lam_3(0,t)c_y(0,t)+{\lam_3}_y(0,t)c(0,t)\right)dt. \nonumber
\end{eqnarray}
}

\paragraph{Constraint 4}

The corresponding term in the inner product is 

$$\int_{0}^T \left(v(1,t)-{s}^{\prime}(t)\right)\lam_4(t) dt,$$
where there is just one derivative of $g$ involved. Integrating by parts we obtain:

\begin{equation}\label{constraint4}
\int_{0}^T \left(\lam_4(t)v(1,t)+{\lam_4}_t(t)s(t)\right)dt - \lam_4(T)s(T)+\lam_4(0)s(0).
\end{equation}

\paragraph{Constraint 5}

Because this term is free of derivatives, there is nothing to do with it, remaining:

\be\label{constraint5}
\int_{0}^T \lam_5(t)v(0,t)dt.
\ee

\paragraph{Constraint 6}

In this case, again, the corresponding term in the inner product remains:

\be\label{constraint6}
\int_{0}^T \lam_6(t)c(1,t)dt.
\ee

\paragraph{Constraint 7}

Even though this term has a derivative, $c_y$, it remains unchanged because function $\lam_7$ depends only on time:

\be\label{constraint7}
\int_{0}^T \lam_7(t)c_y(0,t)dt.
\ee

\paragraph{Constraint 8}

Once more, because of the lack of derivatives the term remains unchanged:

\be\label{constraint8}
\int_{0}^1 \lam_8(y)n(y,0)dy.
\ee

\paragraph{Constraint 9}

In this case, the term is just the product of two real numbers:

\be\label{constraint9}
\lam_9 s(0).
\ee

\paragraph{Obtaining the adjoint equations}

Note that according to equation (\ref{adjoint}), we have to find $\lam$ such that  

\begin{equation}\nonumber
\frac{\partial J}{\partial \phi}g+\left(\frac{\partial E}{\partial \phi}g\right)^*\lambda=0,
\end{equation}
where this equation should be valid for any direction $g\in \mathcal{Y}$.
 
So, putting together equations (\ref{jota2deriv}) with (\ref{constraint1})-(\ref{constraint9}) and choosing the directions conveniently, we get the system of equations which constitutes the adjoint problem (\ref{adj1})-(\ref{adj15}).

\end{document}